\documentclass[amsmath,amssymb,aps,reprint,twocolumn,groupedaddress]{revtex4-2}

\usepackage{graphicx}
\usepackage{bm}
\usepackage[unicode=true,colorlinks=true,citecolor=blue,urlcolor=blue]{hyperref}
\usepackage{braket} 
\usepackage[normalem]{ulem}

\renewcommand{\i}{{i}}

\renewcommand{\d}{\mathrm d}

\begin{document}

\title{Hidden anisotropy controls spin-photon entanglement in a charged quantum dot}

\author{Yuriy Serov}
\email{serovjurij@beam.ioffe.ru}
\author{Aidar Galimov}
\author{Dmitry S.  Smirnov}
\email{smirnov@mail.ioffe.ru}
\author{Maxim Rakhlin}
\author{Nikita Leppenen}
\author{Grigorii Klimko}
\author{Sergey Sorokin}
\author{Irina Sedova}
\author{Daria Berezina}
\author{Yuliya Salii}
\author{Marina Kulagina}
\author{Yuriy Zadiranov}
\author{Sergey Troshkov}
\author{Tatiana V. Shubina}
\author{Alexey Toropov}
\affiliation{Ioffe Institute, 194021 St. Petersburg, Russia}

\date{October 3, 2024}

\begin{abstract}
	Photon entanglement is indispensable for optical quantum technologies. Measurement-based optical quantum computing and all-optical quantum networks rely on multiphoton cluster states consisting of indistinguishable entangled photons. A promising method for creating such cluster states on demand is spin-photon entanglement using the spin of a resident charge carrier in a quantum dot, precessing in a weak external magnetic field. In this work, we show theoretically and experimentally that spin-photon entanglement is strongly affected by the hidden anisotropy of quantum dots, which can arise from mechanical stress, shape anisotropy and even specific crystal structure. In the measurements of time-resolved photoluminescence and cross-polarized second-order photon correlation function in a magnetic field, the anisotropy manifests itself in the spin dynamics and, as a consequence, in the spin-photon concurrence. The measured time-filtered spin-photon Bell state fidelity depends strongly on the excitation polarization and reaches an extremely high value of 94\% at maximum. We specify the magnetic field and excitation polarization directions that maximize spin-photon entanglement and thereby enhance the fidelity of multiphoton entangled states.
\end{abstract}

\maketitle

\section{\label{Introduction}Introduction}
An important resource for the implementation of quantum technologies, including the measurement-based approach of quantum computing~\cite{walther2005, briegel2009, rudolph2017}, is the multiphoton entangled linear cluster state whose nodes are indistinguishable photons~\cite{schon2005}. A promising scheme for the deterministic generation of such states, proposed in general form in Ref.~\cite{schon2005}, was then analyzed in detail theoretically for a quantum dot (QD) photon source by Lindner and Rudolph~\cite{lindner2009}. They considered a singly charged QD in a transverse magnetic field (Voigt geometry). The system generates photons upon optical excitation by pulses of linearly polarized light with a repetition period shorter than the spin precession period of a resident charge carrier in a magnetic field. Polarization entanglement between emitted photons is mediated by a single spin qubit, formed by a charge carrier confined in the QD~\cite{warburton2013}. Each time the QD is excited by a laser pulse, a photon with the polarization entangled with the spin of the carrier is emitted.  By repeating this process over and over, a big cluster of entangled photons with the length limited by the spin lifetime can be created. This approach provides a deterministic process for generating the cluster state, unlike any technique for entangling photons after their emission, which have also been proposed theoretically~\cite{Lee2015_1, Lee2015_2, pilnyak2017} and implemented for a QD-based single-photon source~\cite{istrati2020}.

Recent experiments have demonstrated the feasibility of the Lindner-Rudolph protocol based on the spin of a confined dark exciton~\cite{schwartz2015, schwartz2016} and then on the spin of a hole or electron for positively and negatively charged QDs~\cite{coste2023, cogan2023, su2024}. However, the results were limited to either measuring the spin-photon entanglement only~\cite{coste2023} or to the end-to-end photon generation efficiencies below 2\%~\cite{cogan2023, su2024}. The measured fidelity of the entangled spin-photon state reached only 80\%~\cite{coste2023}, and the fidelity of the multiphoton state did not exceed 71\%~\cite{su2024}.

Several issues have been identified that need to be addressed simultaneously to improve the entanglement robustness and increase the generation rate of the entangled multiphoton state, such as fast spin dephasing due to interactions with surrounding nuclei~\cite{book_Glazov, cogan2018}, low photon collection efficiency without a proper optical microresonator~\cite{senellart2017, lodahl2022}, and polarization distortion caused by splitting or birefringence of the resonator modes~\cite{mehdi2024, leppenen2024}.

However, another critical issue has escaped the attention of researchers, namely the underlying symmetry of QDs. At first glance, for charged QDs, both the ground and excited (trion) states are necessarily degenerate in the absence of a magnetic field according to the Kramers' theorem. Therefore, the anisotropic electron-hole exchange interaction, which spoils the circular polarization of the emission of neutral QDs~\cite{abbarchi2008, seguin2005, mekni2021, kiessling2006}, does not play a role for charged QDs. Although this circumstance masks the possible presence of anisotropy in the QD properties, it has still been shown that the anisotropy affects the polarization of trion photoluminescence in an external magnetic field~\cite{koudinov2004}. The reasons for this are the anisotropic Zeeman splitting of holes and the link between optical selection rules with the crystal axes. Here we show that hidden anisotropy also controls the spin-photon entanglement in charged QDs and, in particular, affects the errors in the Lindner and Rudolph protocol~\cite{lindner2009}.

In this paper, we present a theoretical description of the effects of hidden anisotropy and model experiments on polarization sensitive time statistics of single photons. They uncover the dependence of the spin dynamics and degree of spin-photon entanglement on the directions of excitation polarization and magnetic field in a typical negatively charged single InAs/GaAs QD. We show that the spin-photon concurrence depends on the internal symmetry of the QD, which controls the direction of the hole spin precession axis and, accordingly, its $g$-tensor. As a result, an original approach to boosting entanglement is formulated.

The paper is organized as follows. Section~\ref{Spin-precession-scheme} presents the basic theoretical considerations about the hidden anisotropy in charged QDs. Section~\ref{Experimental} describes the experimental approach to its study.
Section~\ref{Spin_dynamics_in_trion} reports the measured spin dynamics in a single negatively charged QD, successively describing the spin dynamics of the trion (unpaired hole) and the spin dynamics of the resident electron. Section~\ref{Concurrence} describes the concurrence of the spin-photon entangled state, extracted from the measured cross-polarized  photon correlation function $g^{(2)}$. Section~\ref{Discussion} discusses practical steps to optimize multiphoton entanglement. Conclusions are given in Section~\ref{Conclusion}, and detailed descriptions of the experiments and theoretical simulations are given in three appendices.

\begin{figure*}[ht]
  \centering
  \includegraphics[width=0.7\linewidth]{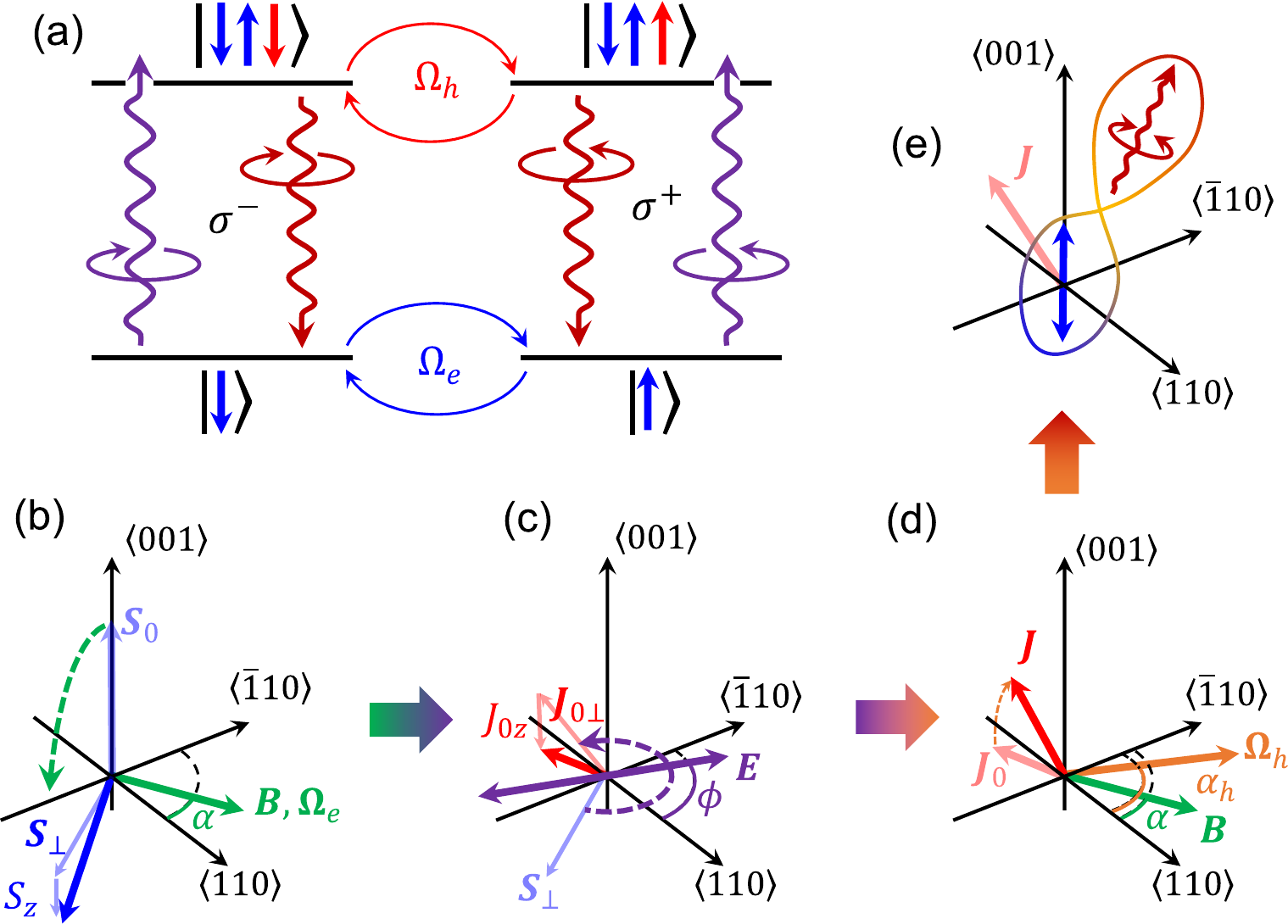}
  \caption{\label{fig_schemes} (a)~Energy levels and optical transitions in a negatively charged QD. (b)---(e) Spin evolution during spin-photon entanglement. (b)~Electron spin precession from $\bm{S}_0$ to $\bm{S}_\perp + \bm{S}_z$ with frequency $\bm{\Omega}_e$ in an external magnetic field $\bm{B}$ [Eq.~\eqref{eq_e_spin_before_exc_simple}]. (c)~Excitation of a trion by linearly polarized light ($\bm{E}$), which initializes the trion pseudospin $\bm{J}_0 = \bm{J}_{0\perp} + \bm{J}_{0z}$ with the $\bm{J}_{0\perp}$ component rotated relative to $\bm{S}_\perp$ [Eq.~\eqref{eq_h_spin_after_exc_simple}]. (d)~Precession of the hole spin from $\bm{J}_0$ to $\bm{J}$ with frequency $\bm{\Omega}_h$ [Eq.~\eqref{eq_h_spin_before_emission_simple}]. (e)~Emission of a photon entangled with the remaining electron according to Eq.~\ref{eq_concurrence_main}.}
\end{figure*}

\section{\label{Spin-precession-scheme}Hidden anisotropy of spin-photon entanglement}

The hidden anisotropy in QDs of arbitrary symmetry can be caused by two features. One of them is the precession of the hole spin in an external magnetic field, since in the general case the direction of the precession axis may not coincide with the direction of the magnetic field. The other is the dependence of the resulting spin of the unpaired excited charge carrier not only on the resident spin before excitation, but also on the polarization of excitation.

Let us describe microscopically the emergence of entanglement between a single photon and an electron spin in a negatively charged QD. We denote the optical axis as $z$ and choose the $x$ and $y$ axes in-plane along $[100]$ and $[010]$ crystallographic directions, respectively. We consider the initial orientation of the electron spin along the optical axis $z$ and the application of an external magnetic field $\bm{B}$ in the Voigt geometry, which leads to the spin precession, as shown in Fig.~\ref{fig_schemes}(a). The precession leads to the electron spin in-plane orientation after time $\pi/(2\Omega_e)$, where $\bm\Omega_e=g_e\mu_B\bm B/\hbar$ is the Larmor frequency with $g_e$ being the electron $g$-factor and $\mu_B$ being the Bohr magneton. To simplify the comparison with experiments, we introduce the angle $\alpha$ between the magnetic field and the $[110]$ axis, see Fig.~\ref{fig_schemes}(b). Then the electron spin components take the form
\begin{equation}
  \label{eq_e_spin_before_exc_simple}
  S_x = \frac12 \sin(\alpha+\pi/4), ~~~~ S_y = -\frac12 \cos(\alpha+\pi/4).
\end{equation}

Next, we consider the instantaneous quasi-resonant excitation, which creates a negatively charged trion in the QD. The spins of the two electrons in the trion are opposite and form a singlet, while the heavy hole can be in one of the two states
\begin{equation}
  \label{eq:hh}
  \frac{\mp\mathcal X-\i\mathcal Y}{\sqrt{2}}\chi_{\pm1/2},  
\end{equation}
which correspond to the projection of the angular momentum $\pm3/2$ to the $z$ axis, respectively. Here $\mathcal X$ and $\mathcal Y$ are the $p$-type orbitals along the corresponding axes, and $\chi_{\pm1/2}$ denote spin-up and spin-down spinors. The optical selection rules illustrated in Fig.~\ref{fig_schemes}(a) imply that the excitation of $\pm3/2$ states is only possible upon absorption of a polarized $\sigma^\pm$  photon in electron $\pm1/2$ spin states. For a linearly polarized $\pi$-pulse, the trion wave function components in the basis~\eqref{eq:hh}, $\psi_{\pm3/2}$, read
\begin{equation}
  \psi_{\pm3/2}=\frac{e_x\mp\i e_y}{\sqrt{2}}\psi_{\pm1/2},
\end{equation}
where $\psi_{\pm1/2}$ are the components of the electron wave function (before excitation) and $\bm e$ is the light unit polarization vector. 
This allows us to relate the trion pseudospin $\bm J$ to the electron spin $\bm S$ before the excitation:
\begin{align}
  \label{eq_h_spin_after_exc_simple}
  J_x &= -S_x\sin 2\phi - S_y \cos 2\phi, \nonumber \\
  J_y &= S_x\cos 2\phi - S_y \sin 2\phi, \\ 
  J_z &= S_z, \nonumber 
\end{align}
where $\phi$ is an angle between $\bm e$ and the $[110]$ axis. This relation is illustrated in Fig.~\ref{fig_schemes}(c). Importantly, the trion spin does not equal simply to the electron spin, but it is rotated in the $(xy)$ plane by an angle $\pi/2 + 2\phi$, which is determined by the light polarization relative to the crystallographic axes.

The trion pseudospin $\bm J$ then begins to precess in the magnetic field. Generally, there can be four linearly independent components $g_{\alpha\beta}^h$ of the hole $g$-tensor in the Zeeman Hamiltonian:
\begin{equation}
  \label{eq_g_h_tensor_in_Hamiltonian}
  \mathcal H_h=\mu_B\sum_{\alpha,\beta=x,y}g_{\alpha\beta}^hJ_\alpha B_\beta.
\end{equation}
Note that in the axially symmetric approximation, the hole spin states with angular momentum $\pm3/2$ cannot be coupled by a magnetic field, because it can change the angular momentum by one only. However, the zinc blende-type crystal lattice itself leads to the fact that the highest possible symmetry of the QD is $D_{2d}$. In this group, there is one symmetry-allowed component of the $g$-tensor: $g^h_{D_{2d}}=(g_{xx}-g_{yy})/2$. Anisotropy of the elastic strain and QD shape with respect to the $[110]$, $[\bar110]$ or $(x,~y)$ axes can further lower the symmetry to $C_{2v}$. The corresponding contributions to the $g$-tensor are $g^h_{C_{2v}}=(g_{xy}-g_{yx})/2$ and $g^h_{C_{2v}'}=(g_{xx}+g_{yy})/2$, respectively~\cite{pikus94,PhysRevB.87.161305,trifonov2021,mauro2024strainengineeringgegesispin}. Finally, if the symmetry is even lower, $C_s$, then a fourth component $g^h_{C_{s}}=(g_{xy}+g_{yx})/2$ arises. As a result, the hole spin precession frequency $\bm\Omega_h=\mu_B\hat{g}\bm B/\hbar$ makes an angle $\alpha_h$ with the $[110]$ axis, which in general differs from $\alpha$, see Fig.~\ref{fig_schemes}(d).

Using Eqs.~\eqref{eq_e_spin_before_exc_simple} and~\eqref{eq_h_spin_after_exc_simple} we obtain the trion pseudospin after precession over time $\tau$
\begin{equation}
  \label{eq_h_spin_before_emission_simple}
  J_z=-\frac{\lambda}{2} \sin \Omega_h \tau
\end{equation}
with the parameter
\begin{equation}
  \label{eq_lambda_definition}
  \lambda=\sin(\alpha_h - \alpha - 2\phi).
\end{equation}
Ultimately, the concurrence of the photon emitted at time $\tau$ and the remaining electron spin [Fig.~\ref{fig_schemes}(e)] is defined by~\cite{leppenen2021, leppenen2024}:
\begin{equation}
  \label{eq_concurrence_main}
  \mathcal C=2J_\perp,
\end{equation}
where $\bm J_\perp=\bm e_xJ_x+\bm e_yJ_y$ with $\bm e_x$ and $\bm e_y$ being the unit vectors along the corresponding axes. Making use of Eq.~\eqref{eq_h_spin_before_emission_simple} we obtain the final result
\begin{equation}
  \label{eq_C}
  \mathcal C=\sqrt{1-\lambda^2 \sin^2\Omega_h \tau}
\end{equation}
(for clarity, we do not account for the spin relaxation here). Note that the electron-photon state is completely entangled ($\mathcal C=1$) for the in-plane hole spin polarization $\bm J=(1/2)\bm J_\perp/J_\perp$, therefore the fidelity in the general case is $\mathcal F=(1+\mathcal C)/2$~\cite{PhysRevA.66.022307,PhysRevA.70.032326}.

Equations~\eqref{eq_concurrence_main} and \eqref{eq_lambda_definition} together clearly demonstrate the hidden anisotropy of spin-photon entanglement: for a finite trion lifetime, the concurrence depends on the directions of light polarization ($\phi$) and in-plane magnetic field ($\alpha$), which determine, in particular, the hole precession axis ($\alpha_h$).

\begin{figure*}[ht]
  \centering
  \includegraphics[width=1\linewidth]{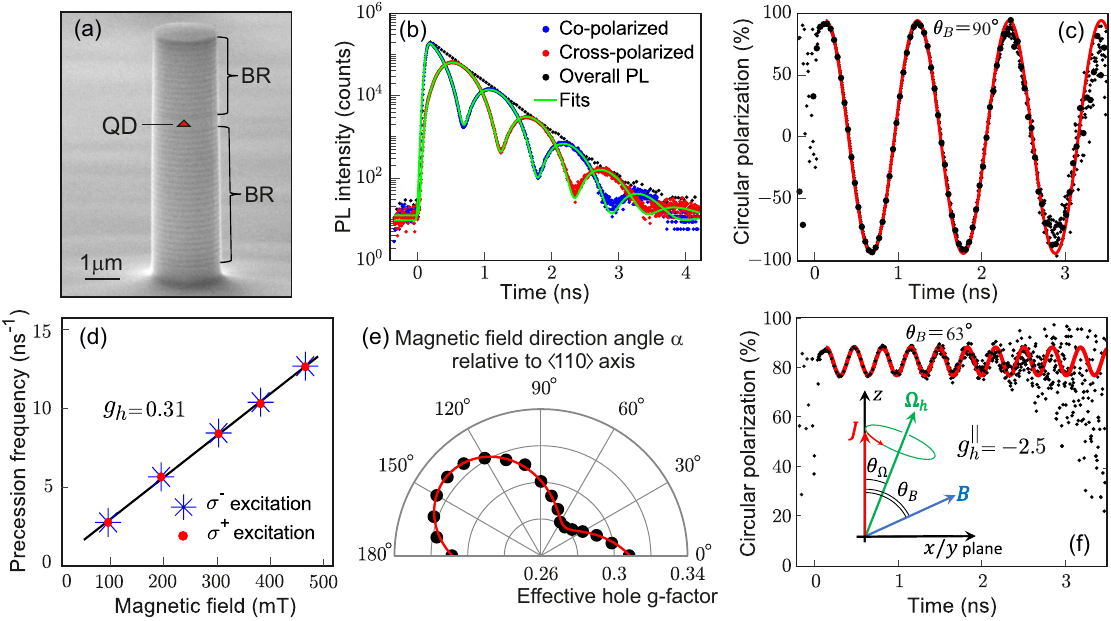}
  \caption{
    (a)~Scanning electron microscope image of a typical micropillar structure in which an InAs/GaAs QD is placed in a $\lambda$-cavity between two Bragg reflectors (BR). (b)~PL decay curves after pulsed $\sigma^-$ LA phonon-assisted trion excitation in co- and cross-polarizations in a transverse magnetic field of 195~mT. The black dots show the sum of the experimental signals, and the lines show the theoretical fit. (c)~The degree of circular polarization of the PL, determined from the dependences shown in figure (b). d)~The hole spin precession frequency as a function of the transverse magnetic field in the [110] direction for $\sigma^+$ and $\sigma^-$ excitations, used to determine the effective $g$-factor. (e)~Anisotropy of the effective $g$-factor over the direction of the magnetic field. (f)~Same as in (c), but in a tilted magnetic field of 183~mT, making an angle $\theta_B=63^\circ$ with the $z$ axis. The inset shows the corresponding hole spin ($\bm J$) precession, taking into account the anisotropic hole $g$-factor.}
  \label{fig_TR}
\end{figure*}

\section{\label{Experimental}Experimentals}

To reveal the predicted hidden anisotropy of spin-photon entanglement with respect to the light polarization and magnetic field directions, we performed a series of measurements on an InAs/GaAs QD located in a post microcavity with distributed Bragg reflectors, optimized for single-photon generation~\cite{rakhlin2023, dryazgov2023}. A scanning electron microscope image of the sample is shown in Fig.~\ref{fig_TR}(a). The structure exhibits high single-photon emission brightness at a wavelength of 919.2~nm and high optical spin orientation of 95\% under circularly polarized excitation. This QD is charged by a single electron, so its energy levels correspond to the diagram in Fig.~\ref{fig_schemes}(a).

We excited the QD quasi-resonantly via a longitudinal acoustic (LA) phonon~\cite{barth2016, cosacchi2019, wang2019, thomas2021} at an energy of $1.7$~meV above the trion resonance. The spin dynamics of a trion, or, equivalently, the photoexcited heavy hole, was studied by polarization- and time-resolved photoluminescence (PL) measurements in a transverse magnetic field. The electron spin dynamics and spin-photon concurrence were studied by measuring the second-order cross-correlation function $g^{(2)}$ between circularly polarized PL components under continuous linearly polarized excitation. The magnetic field applied in these experiments in different directions relatively to the crystallographic axes using permanent magnets was less than 500 mT.

Additional details of the spectroscopic studies of the selected QD and the measurement data are presented in Appendix~\ref{Appendix_methods}. The Appendix includes the PL spectrum, the histogram of the autocorrelation $g^{(2)}$ function, which reveals the single photon purity of 97\%, and the dependencies of the PL line intensity and polarization on the exciting laser detuning from the trion resonance. 

The use of the $g^{(2)}$ function for the analysis of spin-photon entanglement is a new approach. Previously, such functions were used only to study the dynamics of hole spins in positively charged QDs~\cite{coste2022}. 

\section{\label{Spin_dynamics_in_trion}Spin dynamics in a QD}

As shown in Sec.~\ref{Spin-precession-scheme}, the spin-photon entanglement depends on the spin dynamics parameters in both the ground and excited QD states. This section presents experimental measurements for the trion lifetime, the electron $g$-factor, and the hole $g$-tensor.

\subsection{\label{TR_exp} Hole spin dynamics}

The negatively charged trion has a pair of electrons with spin in the singlet state and a heavy hole with unpaired spin, so the trion spin dynamics is determined by the hole spin. It was accessed using polarization- and time-resolved PL measurements in a transverse magnetic field $\bm B$ (Voigt geometry) under pulsed circularly polarized excitation with a repetition rate of 80~MHz. The obtained decay curves of copolarized and cross-polarized PL for $B=195$~mT are shown in Fig.~\ref{fig_TR}(b). The total PL intensity decays monoexponentially and corresponds to the trion lifetime $\tau=370$~ps. This value includes the acceleration of the radiative decay due to the Purcell effect in the microcavity.

The circularly polarized components oscillate in the given magnetic field with a frequency $\Omega_h = 5.4~\text{ns}^{-1}$. The degree of circular polarization of the PL after background subtraction is shown in Fig.~\ref{fig_TR}(c). According to the optical selection rules, it is determined by the heavy hole pseudospin $\bm J$ and is equal to $2J_z$. It oscillates in the magnetic field between $+1$ and $-1$, as expected. We measured the hole spin precession frequency in several magnetic fields for both $\sigma^+$ and $\sigma^-$ polarized excitation. The dependences of the frequency on the transverse magnetic field applied in the $\langle 110\rangle$ direction~\footnote{This denotes either [110] or [$1\bar{1}0$] axis, which we could not distinguish.} are shown in Fig.~\ref{fig_TR}(d). As expected, they are similar for both excitation polarizations. From the linear approximation, the effective $g$-factor of the hole in this direction, defined as $g_{h}^\text{eff} = \hbar\Omega_h/(\mu_B B)$, is $g_h^{\text{eff}}=0.31$.

We performed similar measurements for different in-plane magnetic field directions and found that the effective hole $g$-factor is in the range of $0.28 \div 0.33$. These values are much larger than those for quantum wells~\cite{marie1999}, but are consistent with the previous studies of similar QDs~\cite{trifonov2021,coste2022}.
Note that the angular dependence of the $g$-factor, shown in Fig.~\ref{fig_TR}(e), clearly demonstrates the discrepancy between the crystallographic axes and the directions in which the extreme values of the $g$-factor are observed.

To determine the longitudinal $g$-factor $g_h^\parallel$, which for a heavy hole should be substantially larger than the transverse one~\cite{schwan2011}, measurements were carried out in a tilted magnetic field at different tilt angles $\theta_B$ relative to the optical axis $z$. Figure~\ref{fig_TR}(f) shows much faster oscillations of the PL polarization degree with smaller amplitude in a slightly lower magnetic field compared to panel~(c), but applied at an angle of $\theta_B=63^\circ$. This corresponds to a large hole spin precession vector that is directed almost along the $z$ axis, as shown in the inset.
Fitting the oscillations for several $\theta_B$, we obtained an effective longitudinal $g_h^\parallel$-factor of $-2.5$ (see Appendix~\ref{Appendix_TR_modelling}). This value is in good agreement with that previously measured for a heavy hole in an InAs QD~\cite{belykh2016}. The negative sign is chosen in accordance with previous studies~\cite{cade2006, nakaoka2005}. The large difference between the transverse and longitudinal $g$-factors, characteristic of a heavy hole, unambiguously proves the presence of an electron in the ground state.

The time-resolved PL data presented in Fig.~\ref{fig_TR}(b) can be used to estimate another important parameter, the hole spin relaxation time. While the PL intensity decays over three orders of magnitude, the PL polarization oscillations show no evidence of hole spin relaxation for at least 3~ns in Fig.~\ref{fig_TR}(c). From the fit of the degree of polarization as a function of time, it can be concluded that the hole spin relaxation time exceeds 5.5~ns, which is consistent  with other studies~\cite{coste2022}. The suppressed spin relaxation is related to the weak hyperfine interaction with the lattice nuclei~\cite{book_Glazov, fischer2008, carter2014}, which is known to be responsible for the spin relaxation of various charge complexes in QDs~\cite{PhysRevB.98.121304,cogan2018}. This circumstance clearly indicates the preference of positively charged QDs for the generation of long multiphoton clusters.

\subsection{\label{g2_exp}Electron spin dynamics}

\begin{figure*}[ht]
  \centering
  \includegraphics[width=\linewidth]{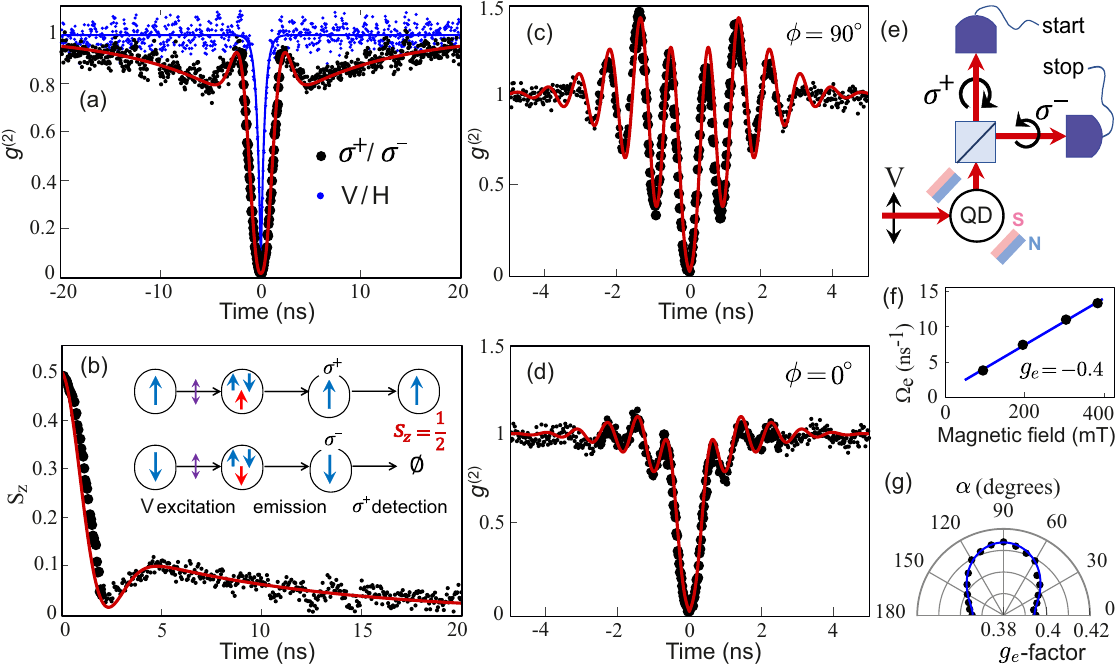}
  \caption{(a)~Cross-correlation $g^{(2)}$ functions of $\sigma^+$ and $\sigma^-$ photons (blue dots) or V and H photons (black dots) under continuous linearly polarized excitation in the absence of a magnetic field. (b)~Electron spin $S_z(t)$ (black dots) determined from the dependence shown in panel (a), and its fit after Eq.~\eqref{eq:MER} (red line). The inset illustrates the process of electron spin initialization by projective spin measurement. (c,d)~Cross-correlation function in a magnetic field of 195~mT applied along the $\langle110\rangle$ direction, measured for excitation polarizations along $\langle1\bar10\rangle$ (c) and $\langle110\rangle$ (d), and their modelling. (e) Start-stop measurement scheme in the circular polarization basis with the application of an external magnetic field. (f) Dependence of the electron spin precession frequency on the magnetic field applied along the $\langle110\rangle$ axis. (g) Anisotropy of the transverse electron $g$-factor.
  }
  \label{fig_g2}
\end{figure*}

To address the electron spin dynamics, we measured the second-order photon cross-correlation function $g^{(2)}(t)$ under linearly polarized continuous wave (CW) excitation using the Hanbury Brown-Twiss setup. The reference cross-correlation measurements in the linear detection polarization basis are shown in Fig.~\ref{fig_g2}(a) with blue dots. The dependence exhibits a typical “single-photon” antibunching dip and can be described as $g^{(2)}(t) = 1 - \exp\left(-t/\tau\right)$ with a trion lifetime $\tau=370$~ps in agreement with the time-resolved PL measurements.

It is important to note that the second-order photon correlation function in the circular detection basis reflects the spin dynamics of the resident electron~\cite{smirnov2017}. For the case of zero magnetic field, the correlation function is shown in Fig.~\ref{fig_g2}(a) by black dots. It has a much wider dip at zero delay time and a nonmonotonous time dependence on its wings. To interpret the results, we note that the detection of the first circularly polarized photon is a projective measurement that initializes the electron spin in the QD~\cite{smirnov2015}, which means that the detection of the first $\sigma^+$ polarized photon is possible for the spin-up electron state only, as shown in the inset in Fig.~\ref{fig_g2}(b). Similarly, the detection of the second $\sigma^-$ polarized photon is possible for the spin-down state of the electron only. Assuming the negligible trion lifetime, we have (for $t>0$)
  \begin{equation}
    \label{eq:g2_Sz}
    g^{(2)}(t)=1-2S_z(t).
  \end{equation}
  The electron spin polarization extracted from the photon correlation function using this relation is shown in Fig.~\ref{fig_g2}(b). It starts from $S_z(0)=1/2$, drops to zero around $t\approx2$~ns, then recovers, and finally decays to zero. All this is in a perfect agreement with the theoretical prediction of Ref.~\cite{merkulov2002}:
\begin{equation}
  \label{eq:MER}
  S_z(t)=\frac{1}{6}\left[1 + 2\left(1 - \frac{2t^2}{{T_2^*}^2}\right)\exp\left(-\frac{t^2}{{T_2^*}^2}\right)\right]\exp\left(-\frac{t}{T_1}\right).
\end{equation}
Here $T_2^*$ and $T_1$ are the electron spin dephasing and longitudinal spin relaxation times, respectively. From the theoretical fit with account for the finite trion lifetime, described in Appendix~\ref{Appendix_g2_no_field}, we obtain $T_2^*=1.6$~ns and $T_1=10$~ns.

We then apply a transverse magnetic field [Fig.~\ref{fig_g2}(e)], and the photon correlation function starts oscillating, as shown in Figs.~\ref{fig_g2}(c,d). The two panels correspond to different linear excitation polarizations, and we will discuss the difference in more detail below.  Here, we note that in both cases the oscillation frequency $\Omega_e$ is the same and increases linearly with increasing magnetic field, as shown in Fig.~\ref{fig_g2}(f). This gives the electron $g$-factor $g_e = -0.4$. Both the in-plane [Fig.~\ref{fig_g2}(g)] and out-of-plane anisotropy of the electron $g$-factor are too small to be reliably determined in our experiments, in contrast to the hole (see Appendix~\ref{Appendix_g2_modelling} for details).

\section{\label{Concurrence}Spin-photon concurrence }

\begin{figure*}[ht]
  \centering
  \includegraphics[width=\linewidth]{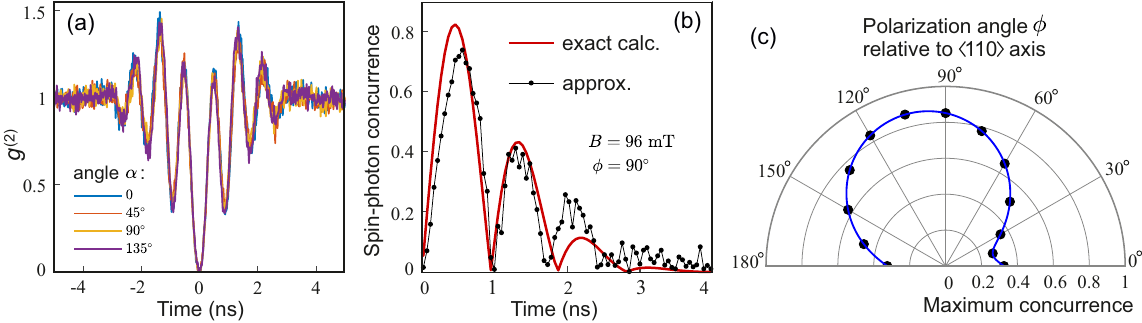}
  \caption{
    (a)~ Cross-correlation function of $\sigma^+$ and $\sigma^-$ photons in a transverse magnetic field $B=195$~mT of different directions specified by the angle $\alpha$ measured relative to the $\langle110\rangle$ axis. The excitation polarization is along $\langle1\bar10\rangle$. (b)~Spin-photon concurrence for photons emitted at a certain time $t$ after spin initialization via projective measurement. Black dots are obtained from experimental values of $g^{(2)}$ measured at $B = 95$~mT using Eq.~\ref{eq_zero_lifetime_concurrence_and_g2}. The red line shows the exact theoretical simulation with experimental parameters. (c)~Maximum spin-photon concurrence at $B = 195$~mT as a function of excitation polarization.
  }
  \label{fig_g2_analysis}
\end{figure*}

The dependence of the photon correlation function in Figs.~\ref{fig_g2}(c,d) on the polarization of the exciting light arises from the spin dynamics of the heavy hole in the trion state, which is described in Sec.~\ref{Spin-precession-scheme}. This dynamics determines the electron-photon entanglement according to Eq.~\eqref{eq_concurrence_main}. To study the dependence on the light polarization and the magnetic field direction, two series of measurements of the correlation function $g^{(2)}$ were carried out: (i)~we fixed the light polarization along the $\langle\bar110\rangle$ axis ($\phi=\pi/2$) and rotated the magnetic field (making an angle $\alpha$ with the $\langle110\rangle$ axis). (ii) We fixed the direction of the magnetic field along the $\langle110\rangle$ axis ($\alpha=0$) and rotated the excitation polarization (making an angle $\phi = 0, 15^\circ, ..., 180^\circ$ with the $\langle110\rangle$ axis).

The measured dependences from series (i), shown in Fig.~\ref{fig_g2_analysis}(a), practically coincide, demonstrating the absence of dependence on $\alpha$. At the same time, the dependence on $\phi$ in series (ii) is clearly expressed, as can be seen in Figs.~\ref{fig_g2}(c,d) (all 13 measurements and their fits are shown in Fig.~\ref{fig_all_g2_over_polar} in Appendix~\ref{Appendix_g2_transverse_field}). Specifically, the amplitude of the oscillations and the width of the dip at zero delay time differ. Such behavior is incompatible with the $S_4$ rotation present in the D$_{2d}$ symmetry group and indicates a decrease in the QD symmetry group to C$ _{2v}$ or lower.

It turns out that all results can be well fitted using the theoretical model introduced in Sec.~\ref{Spin-precession-scheme}, as we describe in Appendix~\ref{Appendix_g2_modelling}. The angles $\alpha$ and $\phi$ enter the modelling only through the parameter $\lambda$ in Eq.~\eqref{eq_lambda_definition}. Therefore, the independence of the form of the $g^{(2)}$ function on $\alpha$ implies that the value of $\lambda$, defined for a fixed polarization by the angle between the hole and electron precession axes $(\alpha_h - \alpha)$, remains almost constant for all in-plane magnetic field directions. From a common fit of all data in series (ii) at $\alpha = 0$, this angle was determined to be equal to $\alpha_h - \alpha \approx -64^\circ$.

The direction of the hole precession axis $\alpha_h$ and the hole effective $g$-factor anisotropy in Fig.~\ref{fig_TR}(e) are both determined by the hole $g$-tensor. From the analysis of the dependencies $\alpha_h(\alpha)$ and $g_h^{\text{eff}}(\alpha)$, presented in Appendix~\ref{Appendix_g_tensor}, we obtain the transverse $g$-tensor in the $(x,y)$ axes:
\begin{equation}
	\label{eq_g_tensor_res}
	\hat g_\perp^h = \begin{pmatrix}
		0.17 & 0.24\\
		-0.28 & 0.14
	\end{pmatrix}.
\end{equation}
In contrast to the studies of QD ensembles, which indicated a diagonalizable form of the transverse hole $g$-factor~\cite{crooker2010, schwan2011}, the measured tensor of an individual QD turns out to be essentially off-diagonal and asymmetric and cannot be diagonalized by any choice of in-plane axes. It corresponds to the following terms in the heavy hole Zeeman Hamiltonian~\footnote{In the modeling the coefficients in the Zeeman Hamiltonian $g^h_{D_{2d}}$, $g^h_{C_{2v}}$, $g^h_{C_{2v}'}$ and $g^h_{C_{s}}$ were used as fitting parameters. So the thousandths in $g^h_{D_{2d}}$ and $g^h_{C_{s}}$ were not direvied from g-tensor, but determined directly.} (Eq.~\eqref{eq_g_h_tensor_in_Hamiltonian}):
\begin{equation}
  \label{eq_g_tensor_components}
  g^h_{D_{2d}} = 0.015,
  ~~
  g^h_{C_{2v}} = 0.26,
  ~~
  g^h_{C_{2v}'} = 0.15,
  ~~
  g^h_{C_{s}} = -0.018.
\end{equation}
Taking into account the weak renormalization due to quantum confinement~\cite{trifonov2021}, the first term agrees well with the value measured in quantum wells~\cite{marie1999}. The last term is small, ang the leading contributions of $g^h_{C_{2v}}$ and $g^h_{C_{2v}'}$ show that the in-plane hole $g$-factor is mainly related to the strain in the QD~\cite{pikus94,linpeng2021}. The dominance of this hole spin mixing mechanism is directly related to the $g^{(2)}$ independence of the magnetic field direction: both contributions $g^h_{C_{2v}}$ and $g^h_{C_{2v}'}$, as well as their arbitrary combination, yield a hole precession axis rotated relative to the magnetic field direction by a fixed angle in the absence of other terms in the Zeeman Hamiltonian.

With this knowledge of the spin dynamics in the QD, the concurrence of the electron spin and the photon polarization was calculated using Eq.~\eqref{eq_concurrence_main}. The detailed calculation procedure is described in Appendix~\ref{Appendix_concurrence}. However, a simple and transparent expression for it can be obtained already from Eq.~\eqref{eq_concurrence_main}, since it describes the hole in-plane spin polarization. For a negligible trion lifetime, it is inherited from the electron spin [see Eq.~\eqref{eq_h_spin_after_exc_simple}], which is directly defined by $g^{(2)}$ [see Eq.~\eqref{eq:g2_Sz}]. Thus, we arrive at
  \begin{equation}
    \mathcal{C}(t) = 
    \frac{1}{\Omega_e}\left|\frac{\d g^{(2)}(t)}{\d t}\right|,
    \label{eq_zero_lifetime_concurrence_and_g2}
  \end{equation}
which uses the relation between the in-plane electron spin polarization and the time derivative of $S_z$ in the magnetic field, neglecting the electron spin relaxation compared to the precession frequency.

This expression directly relates the derivative of the correlation function $g^{(2)}$ to the concurrence after measuring the electron spin precession frequency. We show a comparison of this expression with a detailed modelling in Fig.~\ref{fig_g2_analysis}(b) for a magnetic field $B=96$~mT. This is a typical situation when the precession period ($T=1.9$~ns) is much longer than the trion lifetime $\tau$. It is seen that Eq.~\eqref{eq_zero_lifetime_concurrence_and_g2} directly gives a good estimate for the concurrence. It reaches a maximum around $t\approx T/4$, when the electron spin, initialized along the $z$ axes, makes a quarter turn and lies in the $(xy)$ plane.

Then, from the larger amplitude of oscillations in Fig.~\ref{fig_g2}(c) compared to Fig.~\ref{fig_g2}(d), we can conclude that the concurrence is higher in the former case. The entire dependence of the maximum concurrence on the polarization direction is shown in Fig.~\ref{fig_g2_analysis}(c) for the magnetic field $B = 195$~mT. It changes as much as threefold and reaches a maximum of $\mathcal C = 88\%$ at the polarization direction $\phi\approx103^\circ$, and therefore at $\lambda = 1$, given $\alpha_h - \alpha \approx -64^\circ$. This maximum corresponds to an extremely high time-filtered fidelity of the spin-photon Bell state $\mathcal{F} = 94\%$. Such a pronounced concurrence anisotropy with excellent spin-photon entanglement at the maximum is achieved due to the proximity of the electron and hole effective $g$-factors in a particular QD, as well as due to the special design of the experiment described in Appendix~\ref{Appendix_concurrence_measurements}.

\section{\label{Discussion}Discussion}	

Our studies of the spin-photon entanglement anisotropy for an InAs/GaAs QD showed that the degree of entanglement is very sensitive to the excitation polarization direction, varying by almost a factor of 3 (see Fig.~\ref{fig_g2_analysis}(c)), and under the same conditions is insensitive to the magnetic field direction due to the C$_{2v}$ symmetry of the QD under study, revealed by measuring the transverse hole $g$-tensor. From previous studies of the exciton fine structure in epitaxial QDs~\cite{seguin2005, gaisler2013} it is known that the C$_{2v}$ symmetry of QDs is quite typical. This makes our results important for research in the field of obtaining multiphoton cluster states.

For QDs with other symmetries, the general shape of the $\mathcal{C}(\phi)$ curve will remain unchanged, but the optimal angle $\phi$ will change. This angle for a QD with dominant terms $g^h_{C_{2v}}$ and $g^h_{C_{2v}'}$ in the hole Zeeman Hamiltonian will be determined by the ratio of these terms, being almost independent of the magnetic field direction. If other terms of the Hamiltonian, belonging to the higher symmetry group D$_{2d}$  or the lower symmetry group C$_s$, are sufficiently pronounced, the optimal angle $\phi$ will additionally depend on the magnetic field direction. However, the best entanglement is always achieved for $\lambda=1$ in Eq.~\eqref{eq_lambda_definition}. For example, InAs/GaAs QDs, which allow the generation of entangled photon pairs via a biexciton cascade~\cite{young2005, young2007, skiba2017}, can have a symmetry close to D$_{2d}$. In charged QDs of such a high symmetry, the hole $g$-tensor corresponds to $\alpha_h = -\pi/2-\alpha$.  In this case, the maximum of concurrence is achieved at $\alpha+\phi=\pi/2$. This example shows that excitation with a randomly chosen polarization orientation in a randomly oriented magnetic field can be quite inefficient. 

Since the demonstrated anisotropy arises from the joint spin dynamics of charge carriers of both types, it will manifest itself equally for positive and negative trions with the degree of entanglement anisotropy determined by the products $\Omega_h\tau$ and $\Omega_e\tau$. The magnetic field must be strong enough to rotate the spin by $\pi/2$ faster than it relaxes. Therefore, the effect of hidden anisotropy remains strong for any QDs with a subnanosecond trion lifetime.

\begin{figure}
  \includegraphics[width=0.8\linewidth]{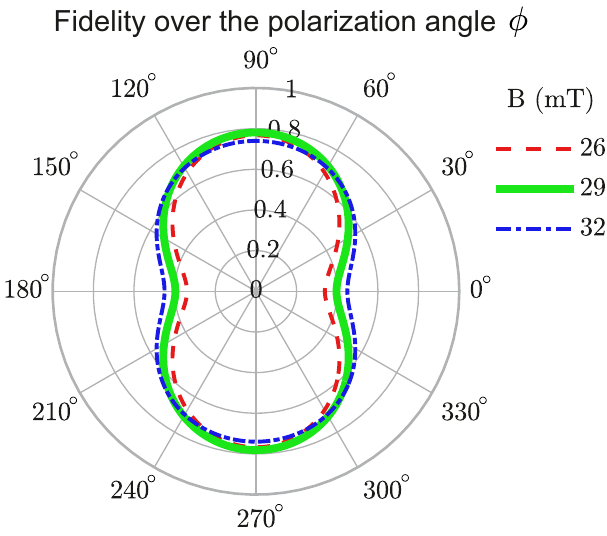}
  \caption{\label{fig_fidelity} Fidelity anisotropy of a 5-photon cluster state generation in a positively charged InAs/GaAs QD for the three values of magnetic field given in the legend.}
\end{figure}

To highlight the importance of anisotropy effects for the generation of multiphoton cluster states, we calculated the fidelity $\mathcal F$ of the $5$-photon cluster state generated in a positively charged QD with the same trion lifetime $\tau = 370$~ps under periodic excitation by linearly polarized quasi-resonant pulses. The positively charged trion was chosen for its longer resident spin relaxation time. The chosen pulse repetition period $5\tau$ was considered optimal in the work~\cite{cogan2023}. We consider the transverse hole $g$-factor of the same scale as in the studied QD, corresponding to one leading term in Zeeman Hamiltonian $g^h_{C_{2v}} = 0.3$. Spin relaxation was not taken into account for simplicity of calculations and interpretation. The calculated fidelity is shown in Fig.~\ref{fig_fidelity} as a function of the light polarization direction relative to $\langle110\rangle$ axis for several magnetic fields, including the optimal $B = 29$~mT, which provides the highest possible entanglement for the simulated system. The best entanglement is achieved when the light is polarized along the $[1\bar10]$ axis, regardless of the magnetic field strength. Notably, a pronounced anisotropy can be seen: the infidelity $1-\mathcal F$ varies for the optimal magnetic field from $0.2$ to $0.6$ --- by a factor of almost~$3$. Thus, regardless of the charged state of the QD, the hidden anisotropy controls an important parameter, the fidelity of the multiphoton cluster states.

\section{\label{Conclusion}Conclusion}

We studied the process of joint spin dynamics of an electron and a photoexcited heavy hole in a single negatively charged InAs/GaAs QD using specially developed correlation techniques. It was found that the observed change in the shape of the correlation function $g^{(2)}$ upon rotation of the excitation polarization is directly related to the anisotropy of spin-photon entanglement. Theoretical analysis showed that this effect arises from the intrinsic anisotropy of the QD, which manifests itself when measuring the tensor of the transverse $g$-factor of a heavy hole. The maximum spin-photon concurrence in typical epitaxial QDs of the $C_{2v}$ symmetry group depends on the direction of excitation polarization, but is almost insensitive to the direction of the magnetic field.

Our results show that when creating spin-photon entanglement and generating multiphoton entangled cluster states using charged QDs, it is necessary to carefully choose the directions of the magnetic field and excitation polarization.

\begin{acknowledgments}
  The work of YS, AG, DSS, MR, GK, SS, IS, DB, YS, MK, YZ, ST, TVS and AT was supported by Rosatom in the framework of the Roadmap for Quantum computing (Contract No. 868-1.3-15/15-2021 dated October 5, 2021 and Contract No. R2152 dated November 19, 2021).
\end{acknowledgments}

\appendix
\section{Methods \label{Appendix_methods}}

\subsection{Fabrication process \label{Appendix_fabrication}}
A planar heterostructure for creating single-photon sources was fabricated by molecular beam epitaxy on a GaAs:Si(001) substrate with a 500 nm thick GaAs buffer layer. The heterostructure contains 25(18) pairs of $\lambda$/4 Al$_{0.9}$Ga$_{0.1}$As/GaAs layers, which form the lower (upper) distributed Bragg reflectors. Between them there is a GaAs $\lambda$-cavity with a thickness of 266 nm, in the center of which an array of InAs QDs with a surface density of 10$^9$ cm$^{-2}$ was formed using the Stranski-Krastanov mode of epitaxial growth.
Reactive ion plasma etching and standard contact photolithography (365 nm, negative photoresist) were used to fabricate regular arrays of 1–3.5 $\mu$m diameter micropillars optimized for single photon emission.
\begin{figure*}[ht]
  \centering
  \includegraphics[width=0.8\linewidth]{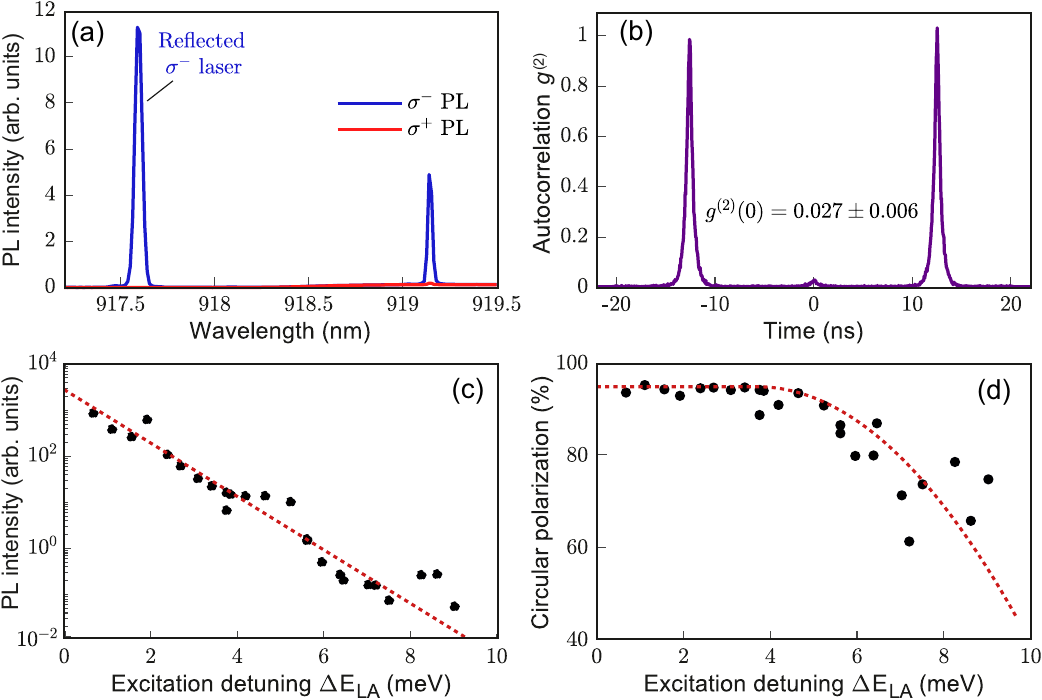}
  \caption{
    (a)~Spectra of circularly polarized $\sigma^-$ and $\sigma^+$ PL components at $919.2$~nm, excited by the $\sigma^-$-polarized laser emission detuned by $\Delta E_\text{LA}=2.4~\text{meV}$. Fully $\sigma^-$-polarized laser light reflected from the sample is also visible. (b)~Autocorrellation $g^{(2)}$ function of unpolarized PL line, excited by the pulsed laser, indicating high degree of single-photon purity. (c)~LA-phonon-assissted excitation efficiency over the laser detuning, which was measured as a time-normalized intensity of PL line, excited by the laser of constant power. (d)~Degree of circular polarization of PL, excited by the $\sigma^-$-polarized laser over the laser detuning. The dashed lines in (c) and (d) are guides to the eye.
  }
  \label{fig_PL_LA-assist}
\end{figure*}

\subsection{Optical measurements \label{Appendix_optical_measurements}}

Microphotoluminescence ($\mu$-PL) spectroscopy was used to study the luminescent properties of the sample. $\mu$-PL spectra were measured at 8 K using a helium flow cryostat (Janis). A wavelength-tunable Ti-sapphire laser was focused on the apex of a selected micropillar using a 0.7 numerical aperture lens. QD luminescence was collected by the same lens and analyzed using a triple grating spectrometer with a cooled CCD camera. Time-resolved and correlation measurements were performed using superconducting single-photon detectors (Scontel) and an SPC-130 (Becker\&Hickl) time-correlated single photon counting system. An external magnetic field of arbitrary direction was generated by two permanent magnets located symmetrically outside the cryostat. Its strength was controlled by the distance between the magnets and measured by a calibrated Hall sensor.

\subsection{Experiment design for concurrence measurements\label{Appendix_concurrence_measurements}}

The spin-photon concurrence was obtained from the analysis of the correlation function (see Appendix~\ref{Appendix_concurrence}) measured between the circularly polarized PL components under CW excitation. The concurrence anisotropy is entirely determined by the decoherence of the QD spin state before photon emission in the statistical ensemble of two-photon detection events that constitute the correlation measurement. This decoherence and its anisotropy were enhanced by the experimental design for clear and unambiguous demonstration of the effect under study.

The decoherence arises from fluctuations in the precession duration of each charge carrier on the trion lifetime scale. Thus, its severity depends on the ratio of the precession periods to the trion lifetime~\cite{lindner2009}. The ratio can be controlled by the strength of the magnetic field (the stronger the magnetic field, the shorter the precession periods and the larger the effect of the fluctuations). In our experiment, we used a magnetic field of 195~mT, whereas values below 100~mT are commonly used to generate cluster states~\cite{coste2023, cogan2023}.

Additionally, we measured the time-filtered concurrence, which characterizes the spin-photon entanglement for a subensemble of photons emitted with a certain delay after spin initialization by a projective measurement. Such a measurement corresponds to a fixed time between the “start” and “stop” photons in the correlation measurement procedure. In this case, despite the fluctuating durations of the electron precession $t_1$ and the hole precession $t_2$, their sum remains strictly fixed in the measured statistical ensemble of events. Such a strong correlation leads to the possibility of either their compensated influence, achieved at $\lambda = 1$, or, conversely, a mutually reinforcing influence, achieved at $\lambda = -1$ with a full range of intermediate possibilities in between. The degree of compensation at $\lambda = 1$ depends on the difference in the precession frequencies of the electron and hole and is therefore controlled by the ratio of the electron and hole g-factors. Thus, in our experiment, decoherence was almost completely compensated at $\lambda = 1$ due to the proximity of the electron and hole effective $g$-factors, revealing a time-filtered concurrence anisotropy.

\subsection{LA-phonon assisted excitation \label{Appendix_LA}}

LA-phonon assisted excitation was used in all experiments. Under such quasi-resonant excitation, a narrow trion emission line with a wavelength of $919.2$~nm was observed in the studied structure.  Figure~\ref{fig_PL_LA-assist}(a) shows the polarized PL components excited by a circularly polarized laser line. These spectra demonstrate the degree of circular polarization of the PL as high as $95\%$,  which is consistent with the long spin relaxation time of the heavy hole $T_h > 5.5~\text{ns}$. The presence of a minor cross-polarized signal can be associated with a non-ideal initial orientation and possible rotation of the light polarization in the microcavity~\cite{arnold2015,hilaire2018,ollivier2020}.
The single-photon nature of the structure emission was confirmed by autocorrelation measurement of the $g^{(2)}$ function in the Hanbury Brown and Twiss setup using pulsed excitation via LA-phonons, as shown in Fig.~\ref{fig_PL_LA-assist}(b). It exhibits high single-photon purity with $g^{(2)}(0)=0.027\pm0.006$.

The excitation using LA phonons is very sensitive to the laser detuning from the resonant transition energy $\Delta E_\text{LA} = E_\text{exc} - E_\text{PL line}$, where $E_\text{ exc}$ is the photon energy of the excitation pulse and $E_\text{PL line}$ is the photon energy of the PL signal. The PL line intensity normalized to the signal acquisition time was measured for constant power excitation with various detunings. Figure~\ref{fig_PL_LA-assist}(c) shows an exponential decay of the intensity with the detuning value $\Delta E_\text{LA}$. However, the degree of PL polarization shown in Fig.~\ref{fig_PL_LA-assist}(d), measured under the same conditions, demonstrates high stability within the detuning variation $\Delta E_\text{LA}<4$~meV.

\section{Modelling details  \label{Appendix_modelling}}

\subsection{Time-resolved PL fitting \label{Appendix_TR_modelling}}

\subsubsection{Dynamics of hole pseudospin and effective $g$-factor \label{Appendix_TR_transverse}}
Time-resolved (TR) measurements of the polarized PL components were carried out to estimate the time dependence of the degree of polarization ($\text{DoP}$) and, therefore, to estimate the dynamics of the heavy-hole pseudospin $J_z = \frac12 \cdot \text{DoP}$. Experimental data of time-dependent PL components intensities $I_+(t)$ and $I_-(t)$ for $\sigma^+$ and $\sigma^-$ polarization detection, respectively, were fitted using the equation:
\begin{equation}
  I_\pm(t) = \left[I_0\cdot \exp\left(-\frac{t}{\tau}\right)\cdot \frac{1 \pm 2\cdot J_z(t)}2\right]*I_\text{instr}(t) + I_{\pm\, \text{bg}},
  \label{eq_TR_fit}
\end{equation}
where $\tau$ is the lifetime of the QD excited state, $I_{\pm\, \text{bg}}$ are the background constant values, $I_\text{instr}(t)$ is the instrumental function of the setup, and $J_z(t)$ reflects the hole pseudospin dynamics. As a result of approximation of the measured shape of the attenuated laser pulse, the instrumental function was adopted as Gaussian with the root-mean-square width $\sigma = 33~\text{ps}$. 

The circularly polarized $\sigma^-$ excitation pulse initializes the hole pseudospin $J_z(0) \approx \frac12$, which then precesses in the external magnetic field. In the Voigt geometry, the function $J_z(t)$ was written as:
\begin{equation}
  J_z(t) = J_{z\, 0}\cdot \cos\left(\Omega_h \cdot t\right),
  \label{eq_J_z_no_relax}
\end{equation}
where the initial hole pseudospin $J_{z\, 0}$ and the precession frequency $\Omega_h$ are the fitting parameters. Equations~\eqref{eq_TR_fit} and \eqref{eq_J_z_no_relax} give an almost perfect fit to the experimental data, as shown in Fig.~\ref{fig_TR}(b), although Eq.~\eqref{eq_J_z_no_relax} does not include the hole spin relaxation. The determined values of $\Omega_h$ in different in-plane magnetic fields $B$ were then used to estimate the hole effective $g$-factor (Figs.~\ref{fig_TR}(d, e)). To estimate the lower bound for the hole spin relaxation time, the $J_z(t)$ functions were modified to include exponential and non-exponential relaxation~\cite{merkulov2002}, and it was found that such a model could not fit the measurements if the heavy hole relaxation time was taken to be less than $5.5~\text{ns}$.

The experimental values of the PL circular polarization for Fig.~\ref{fig_TR}(c) and (f) were determined after fitting the decay curves as:
$$
\text{DoP}(t) = \frac{(I_-(t) - I_{-\, \text{bg}}) - (I_+(t) - I_{+\, \text{bg}})} {(I_-(t) - I_{-\, \text{bg}}) + (I_+(t) - I_{+\, \text{bg}})}.
$$

\subsubsection{Longitudinal hole $g$-factor \label{Appendix_TR_tilted}}
\begin{figure*}[ht]
  \centering
  \includegraphics[width=0.8\linewidth]{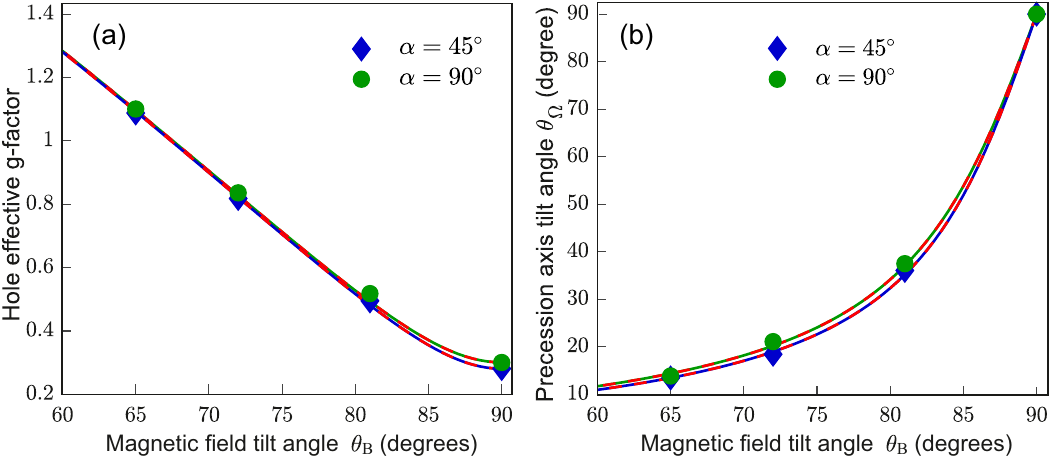}
  \caption{ (a) Dependences of the effective hole $g$-factor $g^h_\text{eff}$ and (b) the hole precession axis tilt angle $\theta_\Omega$ on the tilt angle of the magnetic field $\theta_B$, measured for two directions of the in-plane magnetic field component at angles $\alpha = 45^\circ, 90^\circ$ to the $\langle110\rangle$ crystallographic axis. All lines represent a common fit of all data at $|g^h_{||}| = 2.5$.}
  \label{fig_tilted_TR}
\end{figure*}
In a tilted magnetic field $\bm{B}$ rotated by an angle $\theta_B$ from the normal to the sample surface, the hole precession vector $\bm{\Omega}_h$ rotates from the normal to the surface by another angle $\theta_\Omega$ due to the anisotropy of the $g$-factor of a heavy hole (see the inset in Fig.~\ref{fig_TR}(f)). The hole spin dynamics $J_z(t)$ in the case of an inclined precession axis, neglecting the hole spin relaxation, which is insignificant on the scale of the trion lifetime, is written as:
\begin{equation}
  J_z(t) = J_{z\, 0}\cdot(\cos^2\theta_\Omega + \sin^2\theta_\Omega\cdot \cos(\Omega_h t)).
  \label{eq_J_z_tilted}
\end{equation} 
The precession frequency $\Omega_h$ and the precession axis tilt angle $\theta_\Omega$ for each magnetic field direction were then determined as parameters of the experimental TR data fit using Eqs.~\eqref{eq_TR_fit} and \eqref{eq_J_z_tilted} (see, e.g., Fig.~\ref{fig_TR}(f)). The obtained values of tilt angles and effective $g$-factors corresponding to the extracted precession frequencies $\Omega_h$ are shown in Fig.~\ref{fig_tilted_TR}.   

The precession frequency vector can be expressed in terms of the magnetic field as:
\begin{equation}
  \bm{\Omega}_h = \hat{g}_h\cdot\frac{\mu_B\cdot \bm{B}}{\hbar},
  \label{eq_Omega_h_from_tensor}
\end{equation}
where $\hat{g}_h$ is the hole $g$-tensor. It was found that the experimental data are consistent with the assumption of zero off-diagonal tensor components  $g_{xz}$, $g_{yz}$, $g_{zx}$, $g_{yz}$, which made it possible to describe the tensor in terms of the transverse hole $g$-tensor $\hat g^h_\perp$ and a single value of the longitudinal hole $g$-factor $g_{zz}^h=g^h_{||}$. For an arbitrary direction of the in-plane magnetic field component, specified by the angle $\alpha$ relative to the $\langle110\rangle$ direction, Eq.~\eqref{eq_Omega_h_from_tensor} gives:
\begin{eqnarray}
  \label{eq_Omega_h_tilted}
  \tan(\theta_\Omega) = \frac{g^h_{\text{eff}\,\perp}(\alpha)}{|g^h_{||}|} \cdot \tan(\theta_B), \\
  \Omega_h = g^h_\text{eff}(\theta_B, \alpha) \cdot \frac{\mu_B\cdot B}{\hbar}, \nonumber
\end{eqnarray}
where $g^h_\text{eff}(\theta_B, \alpha)$ is the effective $g$-factor for a certain direction of the magnetic field, and $g^h_{\text{eff}\,\perp}(\alpha)$ is the effective $g$-factor for the corresponding in-plane magnetic field, determined from measurements performed in the Voigt geometry  (Fig.~\ref{fig_TR}(e)). The effective $g$-factor for a tilted field can be expressed using Eq.~\eqref{eq_Omega_h_from_tensor} as:
\begin{equation}	
  g^h_\text{eff}(\theta_B, \alpha) = \sqrt{{g^h_{\text{eff}\,\perp}(\alpha)}^2\cdot \sin^2(\theta_B)+{g^h_{||}}^2\cdot \cos^2(\theta_B)}.
  \label{eq_g_eff_tilted}
\end{equation}
An accurate fit of all experimental data in Fig.~\ref{fig_tilted_TR} can be achieved using Eqs.~\eqref{eq_g_eff_tilted} and \eqref{eq_Omega_h_tilted} with a single fitting parameter $g^h_{||} = -2.5$. Although this fit actually determines the absolute value of the longitudinal $g$-factor, it was assigned a negative sign based on typical values for such QDs~\cite{cade2006, nakaoka2005}.

\subsection{$g^{(2)}$ function fitting \label{Appendix_g2_modelling}}

\subsubsection{General approach \label{Appendix_g2_general}}

In cross-correlation measurements, the $g^{(2)}$ values represent the normalized probability of detecting two orthogonally polarized photons with a time delay $t$ between them. The photon polarization is directly determined by the heavy hole spin at the time of photon emission, which in turn determines the spin of the remaining electron in the QD. Therefore, when approximating the $g^{(2)}(t)$ function, we assume that the electron spin is initialized at time $t=0$ as $S_{z\,0} \approx +\frac12$ upon detection of a $ \sigma^+$-photon, and then the $g^{(2)}(t)$ values are calculated as the population of the spin-down trion state of the QD, normalized to its equilibrium value.

\begin{figure*}[ht]
  \centering
  \includegraphics[width=0.8\linewidth]{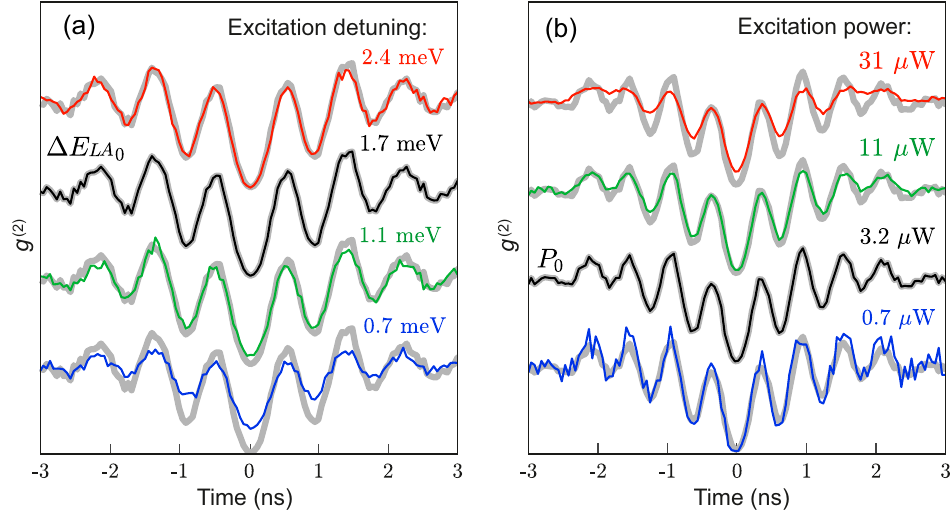}
  \caption{
  	Cross-correlation function of $\sigma^+$ and $\sigma^-$ PL components in a transverse magnetic field of $B=195$~mT (a) or $B=303$~mT (b) for varied parameters of the LA-phonon assisted excitation. Panels (a) and (b) show the variation over the detuning from the QD transition energy and over the excitation power respectively. The gray curves correspond to the excitation power $P_0=3.2~\mu$W and detuning $\Delta E_{LA\,0}=1.7~\text{meV}$, which were used in all other experiments.}
  \label{fig_g2_exc_stability}
\end{figure*}

The model of cross-correlation measurements includes several steps: the electron spin initialization by a projective measurement at time zero, the electron spin dynamics for an arbitrary time $t_1$, the trion state excitation, and then the spin dynamics of the heavy hole (HH) in the trion state for an arbitrary time $t_2$. Analyzing these processes step by step, one can calculate the HH pseudospin projection $J_z$ at $t = t_1+t_2$, from which the cross-correlation function value $g^{(2)}(t)$ is determined by the probability of emitting a $\sigma^-$ photon. Note that the photon emitted by the QD may not be detected, and therefore a two-photon event in the statistical ensemble can include processes with additional steps of the intermediate trion state dynamics. The contribution of these processes should depend on the excitation power, since it is proportional to the probability of emitting an additional photon on the observation timescale. Fig.~\ref{fig_g2_exc_stability} shows the $g^{(2)}$ function measured for different excitation parameters, which clearly demonstrates the shape variation as the excitation power raises significantly. The dependence on the excitation power is consistent with a decrease in the spin precession frequency and an increase in the spin relaxation rate predicted due to the quantum Zeno effect in Ref.~\cite{leppenen2021}. Since for the laser power $P_0=3.2~\mu$W and laser detuning $\Delta E_{LA\,0}=1.7~\text{meV}$ used in all other experiments, the $g^{(2)}$ function is rather stable over the parameters variation, we can conclude, that the contribution of processes with intermediate photon emission is negligible under these conditions and can be ignored. 

The cross-correlation function $g^{(2)}(t)$ at time $t$ is proportional to the population of the trion level with a HH spin down $n_{\downarrow}$, which can be determined as $n_{\downarrow}(t) = n_\text{tr}(t)/2 - J_z (t)$, where the $n_\text{tr} = n_{\uparrow} + n_{\downarrow}$ is the overall excited (trion) state population, and $J_z$ is the HH pseudospin projection, defined as $J_z = \frac12(n_{\uparrow} - n_{\downarrow})$. Here $n_{\uparrow}$ denotes the trion spin-up state population. Then $g^{(2)}(t)$ can be evaluated as 
\begin{equation}
  \label{eq_g2_from_Jz}
  g^{(2)}(t) = \frac{n_{\downarrow}(t)}{{n_\text{tr}}_0/2} = \frac{n_\text{tr}(t) - 2 J_z (t)}{{n_\text{tr}}_0},
\end{equation}
where ${n_\text{tr}}_0$ represents the equilibrium excited state population under CW excitation and allows normalizing the $g^{(2)}$ function to unity, since $J_z$ tends to zero on a large timescale due to spin relaxation. 

In case of weak excitation and, therefore, low equilibrium excited state population, the QD excitation can occur at any time $t_1 < t$ with equal probability density $p_\text{exc}$. After excitation, the excited state population decays exponentially over time $t_2 = t - t_1$, which gives the following expression for the mean value of the excited state population at time $t$ in the statistical ensemble of events: 
\begin{equation}
  \label{eq_n_tr}
  n_\text{tr} (t) = \int\limits_0^t p_\text{exc} \exp\left(- \frac{t_2}\tau \right) \cdot dt_1 = p_\text{exc} \tau \left( 1 - e^{-\frac t \tau} \right),    
\end{equation}
where $\tau$ is the excited state lifetime. Therefore, ${n_\text{tr}}_0 = p_\text{exc} \cdot \tau$ and, thus, in the absence of the spin polarization ($J_z(t) = 0$), Eqs.~\eqref{eq_g2_from_Jz} and \eqref{eq_n_tr} give the well-known expression for the correlation function under the CW excitation: 
\begin{equation}
  g^{(2)}_\text{VH}(t) = 1 - \exp\left(-\frac t \tau\right),
\end{equation}
which was used to fit the reference cross-correlation function measured between the linearly polarized PL components (blue dots in Fig.~\ref{fig_g2}(a)). 

To model the cross-correlation function between the circularly polarized components, the function $J_z(t)$ must also be substituted into Eq.~\eqref{eq_g2_from_Jz}. It can be found similarly to $n_\text{tr}(t)$ by averaging the hole spin before emission over a statistical ensemble of events in which the excitation time $t_1$ fluctuates. Assuming again an equal probability of QD excitation at any time $t_1$ and an exponentially decreasing probability of photon emission after excitation, we obtain a general expression for the mean hole pseudospin at the photon emission time:
\begin{equation}
  \label{eq_mean_J_z_from_J_z_t1_t2}
  J_z(t) = \int\limits_0^t J_z(t_1, t_2) \cdot p_\text{exc} \exp\left(- \frac{t_2}\tau \right) \cdot dt_1,
\end{equation}
where $t_2 = t - t_1$ and $J_z(t_1, t_2)$ is the hole spin at the moment of photon emission $t$, which is determined by the electron spin dynamics over time $t_1$ with subsequent hole excitation and its spin dynamics over time $t_2$. To correctly fit the $g^{(2)}(t)$ functions, the expression for $J_z(t_1, t_2)$ was evaluated separately for the cases of zero magnetic field, applied in-plane magnetic field and tilted magnetic field. Nevertheless, in all cases in the simplified model of the infinitesimal trion lifetime $\tau \to 0$, Eq.~\eqref{eq_mean_J_z_from_J_z_t1_t2} results in the simple relation $J_z(t) = p_\text{exc} \tau J_z(t, 0)$, and the hole spin $z$-component $J_z(t, 0)$ after excitation at time $t$ is equal to the electron spin $S_z(t)$ according to Eq.~\eqref{eq_h_spin_after_exc_simple}. Thus, we obtain:
\begin{equation}
  \label{eq_J_z_simplified_zero_lifetime}
  J_z(t) = p_\text{exc} \tau S_z(t).
\end{equation}
Therefore, in this limit, Eqs.~\eqref{eq_g2_from_Jz}, \eqref{eq_n_tr} and \eqref{eq_mean_J_z_from_J_z_t1_t2} can be simplified to Eq.~\eqref{eq:g2_Sz},
which reflects the main features of the $g^{(2)}$ function.

\subsubsection{Zero magnetic field \label{Appendix_g2_no_field}}

In the case of zero external magnetic field, the only process defining the hole pseudospin $J_z(t_1, t_2)$ after the successive electron and hole spin dynamics over times $t_1$ and $t_2$, respectively, is the electron spin relaxation, because the hole spin relaxation is negligible. Since the hole spin $z$-component inherits the electron spin (Eq.~\eqref{eq_h_spin_after_exc_simple}), the resulting expression is quite simple: 
\begin{equation}
  \label{eq_J_z_t1_t2_no_field}
  J_z(t_1, t_2) = J_z(t_1, 0) = S_z(t_1),
\end{equation}
where the electron spin relaxation function $S_z(t_1)$ is defined by Eq.~\eqref{eq:MER}, but for the smaller relaxation time $t_1 < t$ and possibly non-ideal spin initialization with the initial electron spin $S_{z\,0} \le \frac12$~\cite{merkulov2002,10.1063/1.4984232,book_Glazov}:
\begin{equation}
  \label{eq_S_z_no_field}
  S_z(t)=\frac{S_{z\,0}}{3}\left[1 + 2\left(1 - \frac{2t_1^2}{{T_2^*}^2}\right)\exp\left(-\frac{t_1^2}{{T_2^*}^2}\right)\right]\exp\left(-\frac{t_1}{T_1}\right).
\end{equation}
The perfect fit to the experimental data shown in Fig.~\ref{fig_g2}(a) was obtained using Eqs.~\eqref{eq_g2_from_Jz}, \eqref{eq_n_tr}, \eqref{eq_mean_J_z_from_J_z_t1_t2}, \eqref{eq_J_z_t1_t2_no_field} and \eqref{eq_S_z_no_field}. Note that despite the rather complicated theoretical evaluation of the $g^{(2)}(t)$ values, the only fitting parameters here are the initial electron spin $S_{z\,0}$ and the spin relaxation times $T_1$ and $T_2^*$. Thus, the values $T_2^*=1.6$~ns and $T_1=10$~ns are accurately obtained from the fitting procedure. The fitting curve in Fig.~\ref{fig_g2}(b) was directly evaluated by Eq.~\eqref{eq:MER} and differs slightly.

\subsubsection{Transverse magnetic field \label{Appendix_g2_transverse_field}}

\begin{figure*}[ht]
  \centering
  \includegraphics[width=1.0\linewidth]{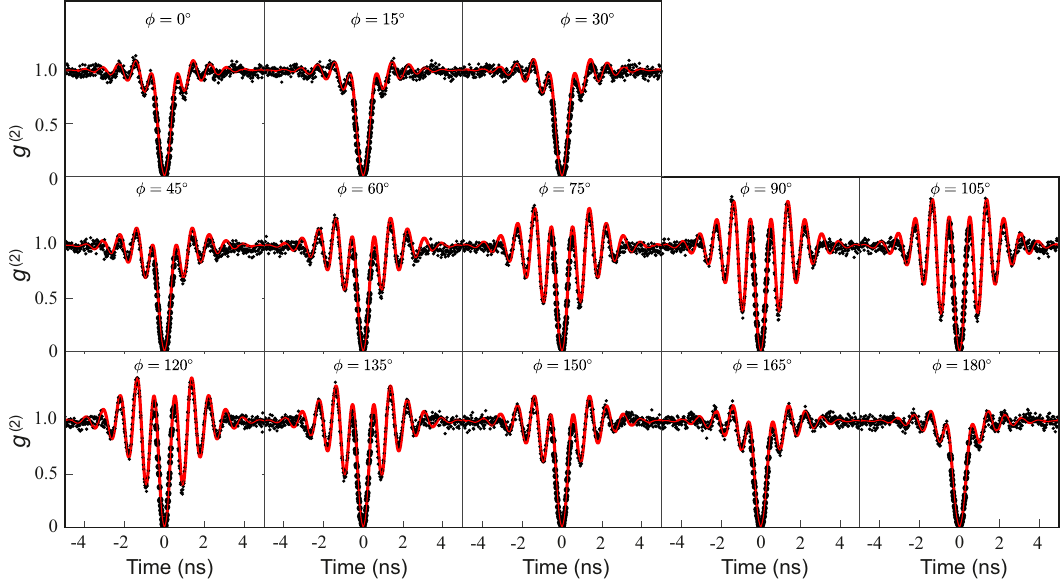}
  \caption{Cross-correlation $g^{(2)}$ function between circularly polarized PL components excited by a linearly polarized laser with polarization direction at different angles $\phi$ from the $\langle110\rangle$ axis, measured in a transverse magnetic field of 195~mT applied along the $\langle110\rangle$ direction.}
  \label{fig_all_g2_over_polar}
\end{figure*}

In a transverse magnetic field, the precession and relaxation of the electron spin during a time period $t_1$ after its initialization by a projective measurement can be described as~\cite{merkulov2002}:
\begin{align}
  \label{eq_e_spin_before_exc}
  S_{\perp\bm\Omega_e} &= S_{z\,0} \sin(\Omega_e t_1) \cdot\exp\left[-\left(\frac{t_1}{T_2^*}\right)^2 - \frac{t_1}{T_1}\right], \nonumber \\ 
  S_z &= S_{z\,0} \cos(\Omega_e t_1) \cdot\exp\left[-\left(\frac{t_1}{T_2^*}\right)^2 - \frac{t_1}{T_1}\right], 
\end{align}
where $S_{\perp\bm\Omega_e}$ denotes the in-plane electron spin component perpendicular to the magnetic field. Then the analysis following the lines of Sec.~\ref{Spin-precession-scheme} yields the component of the hole spin before photon emission for an arbitrary excitation moment~$t_1$:
\begin{multline}
  \label{eq_Jz_t1_t2_Voigt}
  J_z(t_1,t_2) = S_{z\,0}\cdot\exp\left[-\left(\frac{t_1}{T_2^*}\right)^2 - \frac{t_1}{T_1}\right] \cdot \\
  \cdot\left[\cos(\Omega_e t_1)\cdot \cos(\Omega_h t_2) - \lambda \cdot \sin(\Omega_e t_1)\cdot \sin(\Omega_h t_2) \right],
\end{multline}
where the parameter $\lambda$ is given by Eq.~\eqref{eq_lambda_definition} and can be evaluated using experimentally known directions of the magnetic field and polarization. 

The measurements of $g^{(2)}(t)$ in a transverse magnetic field were fitted using Eqs.~\eqref{eq_g2_from_Jz}, \eqref{eq_n_tr}, \eqref{eq_mean_J_z_from_J_z_t1_t2}, \eqref{eq_Jz_t1_t2_Voigt}, and \eqref{eq_lambda_definition} with fitting parameters $S_{z\,0}$, $\Omega_e$ and $\alpha_h$, while $T_1$, $T_2^*$, $\Omega_h$ and $\tau$ are known from the above described measurements. Due to the noisier experimental data in the correlation measurements compared to the TR PL data, the accuracy of the electron precession frequency is significantly lower than the accuracy of the hole precession frequency. For this reason, while we consider the measured in-plane anisotropy of the hole effective $g$-factor at the scale of $\pm8\%$ to be well established (Fig.~\ref{fig_TR}(e)), we completely neglect the measured anisotropy for the electron at the scale of $\pm2.5\%$ (Fig.~\ref{fig_g2}(g)), considering it to be a possible measurement error. We also note that the measured $g^{(2)}$ curves for different magnetic field directions (Fig.~\ref{fig_g2_analysis}(a)) are virtually indistinguishable, which proves the absence of reliable observations of the in-plane anisotropy for the electron $g$-factor.

For fitting the correlation measurements in a series with a fixed magnetic field direction and varying polarization direction, all fitting parameters were kept constant, since they are independent of the polarization direction. These parameters describe simultaneously all the measured data in Fig.~\ref{fig_all_g2_over_polar} solely by changing of $\lambda$ with changing of polarization, as described by Eq.~\eqref{eq_lambda_definition}. Since this equation includes only one fitting parameter $\alpha_h$, which should account for all the difference between these curves, this parameter value $\alpha_h \approx -64^\circ$ is well established from this fit.

\subsubsection{Tilted magnetic field and longitudinal electron $g$-factor  \label{Appendix_g2_tilted_field}}

\begin{figure}
  \includegraphics[width=0.9\linewidth]{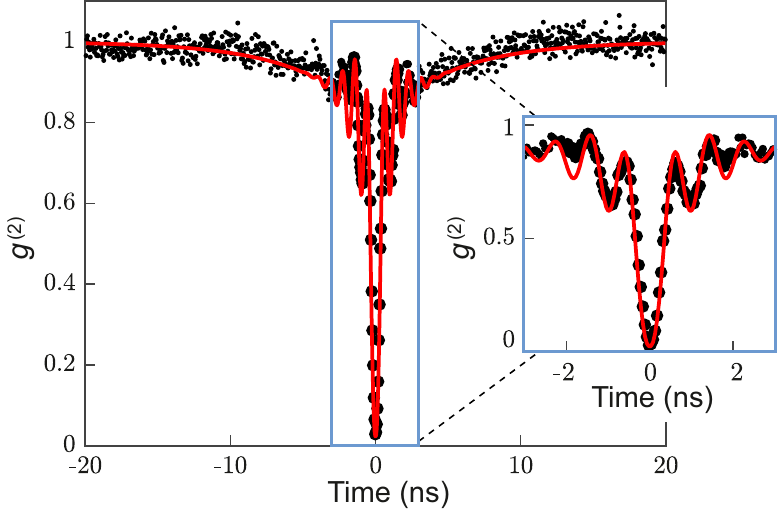}
  \caption{$g^{(2)}$ function in a magnetic field of $183~\text{mT}$, tilted by an angle $\theta_B  = 63^\circ$ from the $z$ axis. The general view shows the long relaxation process of the spin component along the magnetic field direction, while the zoomed inset shows in detail the oscillations at the short time scale.}
  \label{fig_g2_tilted}
\end{figure}

In the case of a tilted magnetic field rotated by an angle $\theta_B$ from the $z$ axis, the sought function $J_z(t_1, t_2)$ was obtained in the same way, assuming an isotropic electron $g$-factor. The obtained function, substituted into Eqs.~\eqref{eq_g2_from_Jz} and~\eqref{eq_mean_J_z_from_J_z_t1_t2}, agrees well with the experimental cross-correlation data, as shown in Fig.~\ref{fig_g2_tilted}. It also clearly demonstrates that the mechanism of fast electron spin relaxation induced by the fluctuating nuclear field is suppressed along the magnetic field direction \cite{merkulov2002}. As a result, the fraction $\cos^2(\theta_B)$ of the initial electron spin undergoes only a slow exponential longitudinal relaxation on the time scale $T_1$.

The isotropic model of the electron $g$-factor was confirmed with the measurement accuracy by analysing the electron precession frequency, which was determined for several tilt angles and allowed us to estimate the values of the electron effective $g$-factor in a tilted magnetic field. These values were found to be almost independent of the magnetic field tilt angle with non-systematic fluctuations in the range of $\pm 3.5\%$, which we attribute to measurement errors. 

\subsection{Spin-photon concurrence evaluation \label{Appendix_concurrence}}

The spin-photon concurrence was determined by calculating two orthogonal components of the in-plane hole pseudospin before the photon emission, $J_{\parallel\bm\Omega_h}(t)$ and $J_{\perp\bm\Omega_h}(t)$, directed along and perpendicular to the hole spin precession axis, respectively. They were determined in the same model as the $J_z$ component, given by Eq.~\eqref{eq_Jz_t1_t2_Voigt}:
\begin{subequations}
  \begin{multline}
    \label{eq_J_x_t1_t2}
    J_{\parallel\bm\Omega_h}(t_1, t_2) = - S_{z\,0}\cdot\exp\left[-\left(\frac{t_1}{T_2^*}\right)^2 - \frac{t_1}{T_1}\right] \cdot \\
    \cdot \sqrt{1-\lambda^2} \cdot \sin(\Omega_e t_1),
  \end{multline}
  \begin{multline}
    \label{eq_J_y_t1_t2}
    J_{\perp\bm\Omega_h}(t_1, t_2) = - S_{z\,0}\cdot\exp\left[-\left(\frac{t_1}{T_2^*}\right)^2 - \frac{t_1}{T_1}\right] \cdot \\
    \cdot\left[\cos(\Omega_e t_1)\cdot \sin(\Omega_h t_2) + \lambda \cdot \sin(\Omega_e t_1)\cdot \cos(\Omega_h t_2) \right],
  \end{multline}
\end{subequations}
followed by averaging over $t_1$ identical to the averaging for the $J_z$ component described by Eq.~\eqref{eq_mean_J_z_from_J_z_t1_t2}. These components were then substituted in Eq.~\eqref{eq_concurrence_main} with an additional normalization by the population of the trion state before emission, $n_\text{tr}(t)$, as follows:
\begin{equation}
  \label{eq_concurrence_detailed}
  \mathcal C(t) = \frac{2 \sqrt{J_{\parallel\bm\Omega_h}^2(t) + J_{\perp\bm\Omega_h}^2(t)}}{n_\text{tr}(t)}.
\end{equation}
The normalization is necessary because the spin-photon concurrence is evaluated assuming that the photon emission occurred at time $t$, which happens with probability proportional to $n_\text{tr}$, given by Eq.~\eqref{eq_n_tr}. 

This calculation was performed for each measured correlation function using the parameters obtained from its fit. Typical results of the concurrence calculation by Eq.~\eqref{eq_concurrence_detailed} using Eqs.~\eqref{eq_J_x_t1_t2}, \eqref{eq_J_y_t1_t2}, \eqref{eq_mean_J_z_from_J_z_t1_t2}, and \eqref{eq_n_tr} are presented in Fig.~\ref{fig_g2_analysis}(b). It was found, that a larger maximum value of the concurrence over time is achieved at a weaker dephasing effect of the excitation moment fluctuation, which corresponds to larger values of the parameter $\lambda$ and a larger amplitude of the $g^{(2)}$ function oscillation. For this reason, the optimal direction of excitation polarization $\phi_\text{max} \approx 103^\circ$, clearly visible in Fig.~\ref{fig_g2_analysis}(c), corresponds to $\lambda = 1$, and therefore it is directly related to the direction of the hole precession axis defined in Sec.~\ref{Appendix_g2_transverse_field}, as $\alpha_h - 2\phi_\text{max} = \pi/2~(\text{mod}~2\pi)$ (see Eq.~\eqref{eq_lambda_definition}, where $\alpha = 0$ due to the choice of the magnetic field direction).

The connection between the cross-correlation function and the spin-photon concurrence becomes clear in the simplified model of an infinitesimally small trion lifetime and a negligible spin relaxation on the timescale of the electron precession period $\tau \ll 1/\Omega_e \ll T_1, T_2^*$. In this case, the normalized hole pseudospin at the emission moment is equal to the electron spin before excitation, since Eq.~\eqref{eq_J_z_simplified_zero_lifetime} can be generalized to all spin components. So Eqs.~\eqref{eq_concurrence_detailed} and \eqref{eq_n_tr} simply give:
\begin{equation}
  \label{eq_concurrence_simplified_from_Sxy}
  \mathcal C(t) = 2 |S_{\perp\bm\Omega_e}(t)|.
\end{equation}
Neglecting the electron spin relaxation, we can derive the in-plane electron spin component from its $z$-component derivative by the equation of spin precession:
\begin{equation}
  \label{eq_S_precession}
  \frac{\d S_z}{\d t} = -\Omega_e \cdot S_{\perp\bm\Omega_e}.
\end{equation}
By combining Eqs.~\eqref{eq_concurrence_simplified_from_Sxy}, \eqref{eq_S_precession}, and \eqref{eq:g2_Sz}, we finally obtain Eq.~\eqref{eq_zero_lifetime_concurrence_and_g2}, which was used for a simplified estimate of the concurrence directly from the $g^{(2)}$ function values in Fig.~\ref{fig_g2_analysis}(b). Note that the assumption $\tau \ll 1/\Omega_e \ll T_1, T_2^*$ directly coincides with the necessary requirements for achieving cluster state generation with acceptable fidelity~\cite{lindner2009}. Therefore, simplified but straightforward Eqs.~\eqref{eq:g2_Sz} and \eqref{eq_zero_lifetime_concurrence_and_g2}, which allow direct estimation of the resident spin dynamics and spin-photon concurrence from correlation measurements, may prove useful in future research on the cluster state generation protocol.

\subsection{Hole $g$-tensor \label{Appendix_g_tensor}}

The hole $g$-tensor defines the hole precession frequency vector for an arbitrary in-plane magnetic field according to Eq.~\eqref{eq_Omega_h_from_tensor}. This equation allows us to evaluate both the effective hole $g$-factor $g_h^\text{eff}$ and the precession axis direction angle $\alpha_h$ for an arbitrary magnetic field direction specified by the angle $\alpha$ relative to the $\langle110\rangle$ axis. Therefore, Eq.~\eqref{eq_Omega_h_from_tensor} was used to fit the $g_h^\text{eff}(\alpha)$ dependence in Fig.~\ref{fig_TR}(e). This fit with four $g$-tensor components as fitting parameters is not well defined and may yield the same fitting curve for different tensors. 

This problem was solved using information about the direction of the hole spin precession axis. From a common fit of the correlation functions in Fig.~\ref{fig_all_g2_over_polar}, it was found that the hole precession axis makes an angle $\alpha_h = -64^\circ$ with the $\langle110\rangle$ direction in the case of a magnetic field oriented along $\langle110\rangle$. This additional requirement on the $g$-tensor was found to be sufficient to make the fit well defined and to obtain the specific form of the hole transverse $g$-tensor, given by Eq.~\eqref{eq_g_tensor_res}.

To check the consistency of all the obtained data and the theoretical model, the dependence $\alpha_h(\alpha)$, completely determined in theory by Eq.~\eqref{eq_Omega_h_from_tensor} with the obtained tensor components, was compared with the experimental data of the cross-correlation function measurements at a fixed polarization direction at $\phi = 90^\circ$ and a varying magnetic field direction. According to Sec.~\ref{Appendix_g2_transverse_field}, the virtually absent changes in the shape of the correlation function with a change in the magnetic field direction, demonstrated in Fig.~\ref{fig_g2_analysis}(a), indicate that the difference $\alpha_h - \alpha$ remains practically constant for all $\alpha$ values, which corresponds to a virtually constant value of the parameter $\lambda$ (Eq.~\eqref{eq_lambda_definition}). A more detailed analysis of the $g^{(2)}$ functions measured in magnetic fields of different directions reveals small changes in the value of $\lambda$ as the magnetic field rotates. The changes $\Delta\lambda(\alpha)$ were determined by fitting the correlation functions according to Eqs.~\eqref{eq_g2_from_Jz}, \eqref{eq_n_tr}, \eqref{eq_mean_J_z_from_J_z_t1_t2}, and \eqref{eq_Jz_t1_t2_Voigt}. Using the known reference value $\alpha_h|_{\alpha = 0} = -64^\circ$, these changes were used to determine the dependence $\alpha_h(\alpha)$ as:
\begin{equation}
  \label{eq_alpha_h_from_lambda}
  \alpha_h(\alpha) = \alpha - 64^\circ + \frac{\d(\alpha_h - \alpha)}{\d\lambda} \Delta\lambda(\alpha),
\end{equation}
where the derivative was taken from Eq.~\eqref{eq_lambda_definition}:
\begin{equation}
  \frac{\d\lambda}{\d(\alpha_h - \alpha)} = \cos(\alpha_h - \alpha - 2\phi) \approx \cos(-64^\circ - 2\phi).
\end{equation}

\begin{figure}
  \includegraphics[width=0.9\linewidth]{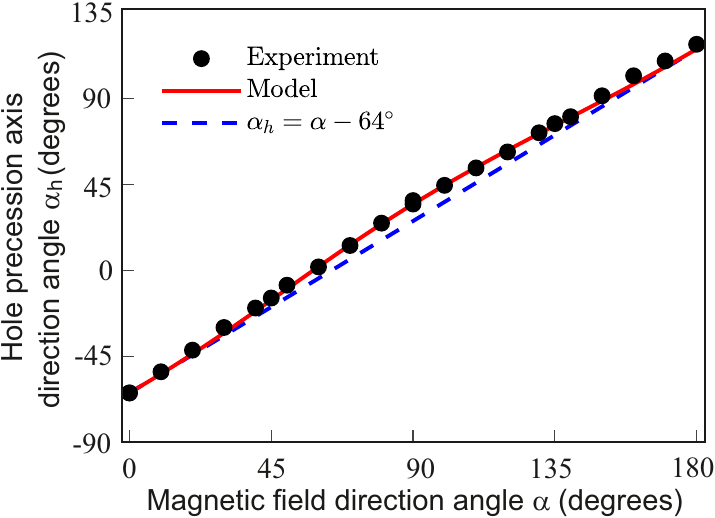}
  \caption{The dependence of the hole precession axis direction on the magnetic field direction, determined from the correlation measurements, which are partially presented in Fig.~\ref{fig_g2_analysis}(a). The experimental values are estimated using Eq.~\eqref{eq_alpha_h_from_lambda}, while the fitting curve corresponds to Eq.~\eqref{eq_Omega_h_from_tensor} with the $g$-tensor given by Eq.~\eqref{eq_g_tensor_res}.}
  \label{fig_alpha_h_over_alpha}
\end{figure}

This approach was used because the changes $\Delta\lambda$ can be reliably determined from the fits of two $g^{(2)}$ functions with other fitting parameters fixed. The value of $\lambda$ itself cannot be estimated for each individual $g^{(2)}$ curve with sufficient accuracy, because it affects the curve shape together with other fitting parameters.

The dependence $\alpha_h(\alpha)$ thus obtained is presented in Fig.~\ref{fig_alpha_h_over_alpha}. It is clearly seen that it is almost linear, which corresponds to the relation $\alpha_h - \alpha \approx \text{const}$, visualized by the blue dashed line. However, it was found that the experimentally observed deviation from this relation is perfectly described by the obtained hole $g$-tensor.

\section{Fidelity of a cluster state  \label{Appendix_fidelity}}
The importance of taking into account the spin-photon entanglement anisotropy is illustrated by the dependence shown in Fig.~\ref{fig_fidelity}. It is calculated for the protocol described in Ref.~\cite{lindner2009}. In particular, we assume that the $\pi$ pulses arrive at time intervals $T_R$, and the first pulse arrives when the electron spin is polarized along the $x$ axis. Moreover, we assume that the trion excited by the $n$-th pulse lives for a time $\tau_n$.

Then the wave functions immediately after the emission of $n=1,2,\ldots$ photons have the form
\begin{subequations}
  \label{eq:Psis}
  \begin{equation}
    \Psi_1\propto(c_1-s_1)\uparrow R+(c_1+s_1)\downarrow L,
  \end{equation}
  \begin{multline}
    \Psi_2\propto(c_1-s_1)\left\{\left[\left(\tilde c_1+\tilde s_1\right)c_2-(\tilde c_1-\tilde s_1)s_2\right]\uparrow R
    \right.\\\left.
      +\left[\left(\tilde c_1+\tilde s_1\right)+\left(\tilde c_1-\tilde s_1\right)c_2\right]\downarrow L\right\}R\\+(c_1+s_1)\left\{\left[\left(\tilde c_1+\tilde s_1\right)c_2-\left(\tilde c_1-\tilde s_1\right)s_2\right]\downarrow L
    \right.\\\left.
      -\left[\left(\tilde c_1\tilde s_1\right)c_2+\left(\tilde c_1+\tilde s_1\right)s_2\right]\uparrow R\right\}L,
  \end{multline}
\end{subequations}
and so on. Here $c_n=\cos(\Omega_h\tau_n/2)$, $s_n=\sin(\Omega_h\tau_n/2)$, $\tilde c_n=\cos(\Omega_e\tau_n/2)$, $\tilde s_n=\sin(\Omega_e\tau_n/2)$, where $\Omega_e$ and $\Omega_h$ are the spin precession frequencies in the ground and trion states, respectively; $\uparrow/\downarrow$ denote the electron spin-up/down states, $R$ ($L$) denote $\sigma^+$ ($\sigma^-$) photons in the order of their emission from right to left. The phases of the states are chosen according to Ref.~\cite{lindner2009} and we consider $T_R\Omega_e=\pi/2$. The spin precession frequencies are read as $\Omega_{e,h}=g_{e,h}\mu_BB/\hbar$, where $g_e$ and $g_h$ are the electron and hole $g$-factors, respectively, $\mu_B$ is the Bohr magneton, and $B$ is the external magnetic field.

The ideal cluster states Cl$_n$ can be obtained from Eqs.~\eqref{eq:Psis} by setting $\tau_n=0$. The fidelity of the final state can be calculated as
\begin{equation}
  \label{eq:Fn}
  \mathcal F_n=\int\limits_0^\infty \d\tau_1\ldots\d\tau_n\frac{\exp(-\tau_1/\tau\ldots-\tau_n/\tau)}{\tau^n}f_n,
\end{equation}
where $\tau$ is the trion lifetime and
\begin{equation}
  f_n=\left<{\text{Cl}}_n\middle|\Psi_n\right>^2.
\end{equation}
From Eqs.~\eqref{eq:Psis} one can see that
\begin{subequations}
  \begin{equation}
    f_1=c_1^2,
  \end{equation}
  \begin{equation}
    f_2=c_1^2(\tilde c_1c_2+\tilde s_1s_2)^2,
  \end{equation}
  \begin{equation}
    f_n=f_{n-1}(\tilde c_{n-1}c_n+\tilde s_{n-1}s_n)^2.
  \end{equation}
\end{subequations}
The integration in Eq.~\eqref{eq:Fn} can be done analytically, but the result is cumbersome. The calculation result for $n=5$ in shown in Fig.~\ref{fig_fidelity} for different magnetic fields indicated in the legend.

\bibliography{literature}
\end{document}